\documentclass[10pt,twoside]{classe-cmb}
\usepackage{amssymb,amsbsy,amsmath,amsfonts,amssymb,amscd}
\usepackage{latexsym,euscript,exscale,epic,eepic,epsfig}
\usepackage[english,francais]{babel}
\usepackage{times}
\usepackage[latin1]{inputenc}
\newcommand*{\beq}{\begin{equation}}
\newcommand*{\eeq}{\end{equation}}
\newcommand*{\beqa}{\begin{eqnarray}}
\newcommand*{\eeqa}{\end{eqnarray}}
\newcommand*{\ba}{\begin{array}}
\newcommand*{\ea}{\end{array}}

\newcommand{\boom}{{\sc Boomerang} }

\newtheorem{e-proposition}[theorem]{Proposition}

\newtheorem{e-definition}[theorem]{Definition\rm}

\ComParit{S1296-2147}
\PIT{FLA}
\PXHY{????}
\Add{?}
\Volume{0}
\Year{2003}
\FirstPage{1}
\LastPage{??}
\AuteurCourant{Delabrouille, Kaplan, Piat, Rosset}
\TitreCourant{Polarization experiments} 
\Journal
\Rubrique{Rubrique}{Heading}
\SousRubrique{Sous-rubrique}{Sub-Heading}
\PresentePar{Jacques}{DELABROUILLE}
\Recu{blank}{}{}
\begin{document}
\bibliographystyle{unsrt}
\selectlanguage{english}
\TitleOfDossier{
}
\title{%
Polarization experiments
}
\author{%
J. Delabrouille~$^{\text{a}}$,\ \
J. Kaplan~$^{\text{b}}$,\ \
M. Piat~$^{\text{c}}$
C. Rosset~$^{\text{d}}$
}
\address{%
\begin{itemize}\labelsep=2mm\leftskip=-5mm
\item[$^{\text{a}}$]
PCC Coll\`ege de France, Paris,
E-mail: delabrouille@cdf.in2p3.fr
\item[$^{\text{b}}$]
PCC Coll\`ege de France, Paris,
E-mail: kaplan@cdf.in2p3.fr
\item[$^{\text{c}}$]
PCC Coll\`ege de France, Paris,
E-mail: piat@cdf.in2p3.fr
\item[$^{\text{d}}$]
PCC Coll\`ege de France, Paris,
E-mail: rosset@cdf.in2p3.fr
\end{itemize}
}
\maketitle
\thispagestyle{empty}
\begin{Abstract}{%
Possible instrumental set--ups for the measurement of CMB polarization are reviewed in this article. We discuss existing and planned instruments, putting special emphasis on observational, instrumental, and data processing issues for the detection of very low polarization signals of prime cosmological interest. A short prospective summary is included.

{\bf\large R\'esum\'e.}  
Nous pr\'esentons les dispositifs exp\'erimentaux pour la mesure des anisotropies de polarisation du fond de rayonnement cosmologique et les principaux instruments existant et en cours de r\'ealisation. Nous insistons plus particuli\`erement sur les aspects relevant de l'instrumentation, des observations, et du traitement des donn\'ees pour la d\'etection des signaux extr\^emement faibles qui v\'ehiculent l'information la plus interessante pour la cosmologie. Nous terminons sur une courte discussion prospective.
}\end{Abstract} 
\par\medskip\centerline{\rule{2cm}{0.2mm}}\medskip
\setcounter{section}{0}
\selectlanguage{english}

\section{Introduction} 
After the success of Cosmic Microwave Background (CMB) anisotropy measurements in the past decade, a large fraction of the CMB experimental community now turns towards building experiments for measuring the small fraction of polarized emission in the CMB. 

Polarization signals are much smaller in amplitude than temperature anisotropies. Polarization fluctuations from scalar modes are about one tenth of those of CMB anisotropies on small scales, and still much lower (relatively) on large scales. The most interesting polarization signature, that of gravity waves from inflation, is even smaller by yet an order of magnitude, possibly more. Hence, very sensitive instruments are needed to measure the polarization signals of the CMB, about 100 times more sensitive than present--day experiments.

The process of CMB polarization generation, and motivations for CMB polarisation measurements, are reviewed in \cite{kaplanCRAS}.
In the present review paper, we summarize the main issues, the technological developments, and the experimental program for CMB polarization measurements.

\section{Measuring polarization}

The polarization state of an incoming transverse electromagnetic wave is fully described, in a given reference frame, by the four Stokes parameters $I$, $Q$, $U$, and $V$ defined as:

\begin{eqnarray}
I = \langle |E_x|^2 \rangle + \langle |E_y|^2 \rangle \\
Q = \langle |E_x|^2 \rangle - \langle |E_y|^2 \rangle \\
U = \langle E_xE_y^\dagger \rangle + \langle E_yE_x^\dagger \rangle \\
V = i(\langle E_xE_y^\dagger \rangle - \langle E_yE_x^\dagger)\rangle
\end{eqnarray}

The measurement of all Stokes parameters at a given pixel requires a combination of the measurements of 4 integrated quantities at least.

For CMB polarization measurements, one may decide to measure only the $I$, $Q$ and $U$ Stokes parameters (total intensity and linear polarization), and disregard $V$. In this case, at least three measurements are needed. One can use for this purpose total--power detectors sensitive only to one direction of polarization, as discussed by Couchot and collaborators \cite{1999A&AS..135..579C}. This kind of linear polarimeter set--up has been selected for the measurement of polarization with the B2K version of \boom \cite{montroy} and with the Planck mission \cite{Planck-polar}.

Such measurements, however, disregard an important tool for checking the consistency and robustness of the observations, as it does not permit verifying that the $V$ Stokes parameter behaves as expected from CMB theoretical predictions. The ultimate CMB polarization experiment must have the ability to measure all Stokes parameters of the incoming radiation.

\subsection{What type of experiment?}

Several options are open for CMB polarization measurements: imaging or interferometric observation scheme; ground based, balloon--borne, or space--borne experiment; bolometric or radio detectors; how many frequency channels; what technology...

For CMB anisotropies, the most successful choices have been 
\begin{itemize}
\item radio detectors in space: the DMR instrument on \emph{COBE} \cite{1990ApJ...360..685S}, the Wilkinson Microwave Anisotropy Probe WMAP\cite{2003ApJ...583....1B}, for the lower $\ell$ range, 
\item balloon--borne bolometer instruments: MAXIMA \cite{2000ApJ...545L...5H}, in the middle $\ell$ range, Arche\-ops \cite{2002APh....17..101B}, the Boomerang experiment \cite{2002ApJS..138..315P},  
\item ground-based interferometers: DASI \cite{2002ApJ...568...28L}, the Cosmic Background Imager CBI \cite{2003ApJ...591..556P}, the Very Small Array VSA \cite{2003MNRAS.341L..23G}) at high $\ell$, with very recently the notable exception of ACBAR \cite{kuo-acbar}, which provided very sensitive measurements of the CMB power spectrum at high $\ell$ using ground--based bolometer arrays at the South Pole.
\end{itemize}

The reasons for this are readily understood. For low spectral resolution experiments, bolometers are more sensitive than radio detectors. However, observations being typically carried out at frequencies above 100 GHz, bolometric experiments have suffered badly from atmospheric emission, unless they have observed from a very good site (high altitude, very dry). Bolometers also need to be cooled down to cryogenic temperatures (300 mK or, even better, 100 mK), which makes their use in orbit quite challenging. The technology is getting mature just for the time for (and under the impulse of) the Planck project, a sensitive space mission to be launched in 2007. Hence, bolometers have been competitive so far only in sub-orbital experiments, in situations with little atmospheric emission, and especially on balloons (in spite of the limited integration time, in the range of few hours up to few days). For such balloon experiments, payload mass constraints limit the size of the gondola, and hence of the primary mirror, so that resolution is typically limited to about ten arcminutes. Sky coverage is limited to few tens of per cents at best. Such experimental set--ups gives unchallenged access to intermediate $\ell$ at relatively low cost and within reasonably short time schedules (as compared to space missions).

Radiometers, an older but more mature technology for CMB measurements, operate at lower frequency, and hence provide lower resolution at the diffraction limit for the same size of the optics. Because their use in space was not as challenging as that of bolometers, they have already been flown on two CMB space missions (\emph{COBE} and WMAP), which provide full sky coverage with long integration time, and hence the best present--day measurements of temperature anisotropies in the low $\ell$ range.

Interferometers permit to achieve high resolution without need for very large antennas or telescopes. Typically, a set of 20 intermediate--size half--meter class dishes operating at 30 GHz, configured in close packed array format, probe scales from few arcminutes to about a degree. Arcsecond resolutions can be reached with long baseline configurations. Ground--based interferometers, because of the modulation due to the tracking of celestial sources, are efficient at rejecting atmospheric emission, the fringes of which are distinct from those on the sky. Hence, ground--based interferometers provided the first measurements in the high $\ell$ range.

Prior to all results, reaching a conclusion on what would be the most successful technological and observational choices was far from obvious. For instance, several ground--based bolometric observations with the SuZie experiment \cite{1997ApJ...479...17H}, at the MITO telescope in the italian alps \cite{1999NewAR..43..297D}, with Diabolo \cite{2000A&AS..141..523B,2002ecmw.conf..116D}, have been made in the hope of reaching competitive sensitivities. While quite successful on SZ--effect observations towards several clusters of galaxies \cite{1999ApJ...519L.115P,2000ApJ...538..505M,2002ApJ...574L.119D}, most of these experiments have yielded only upper--limits (or marginal detections) on CMB anisotropies \cite{1997ApJ...484..523C}. 

Similarly, no balloon-borne radiometer instrument has produced any major measurement of the CMB anisotropy spectrum either, as the moderate improvement of sensitivity obtained from flying radiometers on a balloon did not compensate for the huge loss of observation time as compared with ground--based radiometer instruments. In addition, the competition with satellites was in favor of the space--borne instruments. 

We are now in the situation of making the same difficult strategical choice for polarization measurements. This bet has significant impact on further technological developments. A simple transposition of the conclusions reached for CMB anisotropies does not hold \emph{a priori}, as the polarized and total emission of the atmosphere and the foregrounds do not necessarily peak at the same scales and frequencies. In addition, polarization experiments are not sensitive to the same systematics as total intensity ones.

\subsection{Major experimental issues}

\subsubsection{Sensitivity}

One of the key issues for measuring CMB polarization is the extreme sensitivity required, several orders of magnitude better than that required to measure CMB anisotropies.

The best present detectors in sensitive balloon-borne CMB experiments are already photon noise limited. Therefore, reaching sensitivities better by two orders of magnitude cannot be achieved only by reducing detector noise. The gain in sensitivity requires about 10,000 times more detectors or 10,000 times more integration time, or a compromise between the two.

The need to integrate for long observation times is in favor of ground--based or space--borne instruments, although very long duration (few months) ballooning may prove competitive if actually used. The need to integrate many detectors is in favor of bolometer arrays in which many detectors can be produced at the same time by photolithography. A critical question is whether the atmosphere will be as bad for ground based polarization measurement with bolometers as it has been for bolometric anisotropy measurements. The development of the infrastructure needed to put observing sites in the best locations in the world, as the Franco--Italian station at dome C in Antarctica, or the Atacama desert in Chile, may help. It is possible (and largely argued in the community) that atmospheric emission -- expected to be essentially unpolarized at millimeter wavelengths, should be less of a problem for polarization measurements than it has been for anisotropies.

\subsubsection{Foregrounds}

For CMB anisotropy mapping, it has been possible to select the wavelength(s) of observation in a window of the electromagnetic spectrum where CMB dominates over foregrounds, at least in some clean regions of the sky.
In addition, the near-independance of most emission laws on sky coordinates allows for subtraction of foreground contaminants using a proper linear combination of measurements at a few nearby wavelengths. If foreground emissions had been ten or hundred times worse relatively, the detection of CMB anisotropies would have taken much longer!

The situation is unclear for polarization measurements. The situation in terms of emission amplitudes may be somewhat worse than for temperature, as the CMB is only a few per cent polarized, while synchrotron emission can be up to 70\% polarized, and dust emission has been measured by Archeops to be up to 10--15 \% polarized in some regions \cite{archeops-polar}. 

For a complicated polarized foreground sky emission, disentangling the various contributors may require a broader spectral coverage, narrower frequency bands, and more analysis efforts than is common practice in CMB anisotropy observations.

In any case, sensitive multifrequency observations are required for investigating further the situation as far as foregrounds (including atmospheric effects) are concerned. Current experiments already provide useful measurements of foreground polarization properties, but these measurements do not provide us yet with all the information needed to decide which observational strategy is the best to reject optimally polarized foreground contamination. 

\subsubsection{Systematic effects} 

One of the key issues for polarization measurements is that actual detector measurements are linear combinations of the Stokes parameters $I$, $Q$, $U$ and $V$
(or of $T$, $E$, $B$, $V$), and not of single Stokes parameters directly. Recovering the Stokes parameters requires inverting a (linear, at least to first order) system and separate contributions of very different amplitudes. For the CMB for instance, $T \gg E \gg B \gg V$. Imperfect knowledge of the system results in inversion errors and leakage of large $I$ into smaller polarization parameters.

A class of systematic effects arises from the problem of the imperfect modeling of the instrument (and hence of the data). In polarization experiments which involve differencing large signals to measure a tiny fractional polarization, the problem of matching the responses of the two detectors used is of prime importance. Spectral bands, beam shapes, amplifier gains, time--responses, must all be matched as well as possible. WMAP, Planck, B2K and other experiments will have to cope with some or all of these potential systematic effects.

Very general to CMB measurements (and also quite a worry for CMB polarization) is the problem of stray light. Radiation pick--up from the sky in the far side lobes of the antenna pattern, as well as from thermally unstable instrumental parts, has always worried CMB experimenters. For polarization, because of the extreme sensitivity needed, the problem is even more acute. 

Sidelobe levels are greatly reduced by appropriate choice of the optics coupling the detectors to the sky. Corrugated horns, Lyot stops, and oversized mirrors are the typical tools to sidelobe minimisation. For some polarization experiment designs, in particular integrated bolometer arrays located directly in the focal plane of a telescope, special care may be needed to avoid the problem of stray radiation.

\section{Detector technology for CMB polarization measurements}

The scientific case for CMB polarization measurements has motivated a large fraction of the CMB instrumentalists towards the development of the technology needed to measure very tiny CMB polarization signals.

\subsection{Correlators}

One of the simplest ways to access polarization Stokes parameters, at least in principle, is to correlate directly signals proportional to the electric field. Such correlators are vastly used in CMB polarization experiments, in AMIBA \cite{2000IAUS..201E..31L}, DASI, PIQUE/CAPMAP, SPOrt \cite{2003SPIE.4843..305C}, for instance. 

Such correlations can be computed for single--horn signals, as in PIQUE, CAPMAP or SPOrt, or for multiple horn signals in an interferometer--type experiment, as DASI and AMIBA. The limitation of the interferometer set--up is the number of correlators needed for a large focal plane array of receivers.

\subsection{Polarization Sensitive Bolometers}

Jones and collaborators \cite{jones-psb} have developed a bolometric detector that is intrinsically
sensitive to linear polarization and is optimized for making
measurements of the polarization of the cosmic microwave
background radiation. The receiver consists of a pair of
co-located silicon nitride micromesh absorbers which couple
anisotropically to linearly polarized radiation through a
corrugated waveguide structure. This system allows simultaneous
background limited measurements of the Stokes $I$ and $Q$
parameters over $\sim 30$\% bandwidths at frequencies from $\sim
60$ to 600 GHz. Since both linear polarizations traverse identical
optical paths from the sky to the point of detection, the
susceptibility to systematic effects is reduced. The amount of
uncorrelated noise between the two polarization senses is limited
to the quantum limit of thermal and photon shot noise, while
drifts in the relative responsivity to orthogonal polarizations
are limited to the effect of non--uniformity in the thin film
deposition of the leads and the intrinsic thermistor properties.
Devices using NTD Ge thermistors have achieved NEPs of $2 \cdot
10^{-17} ~ \mathrm{W}/\sqrt{\mathrm{Hz}}$ with a $1/f$ knee below
50 mHz at a base temperature of 270 mK.

Such detectors have been flown successfully on the Boomerang experiment (B2K). However, as these are individual detectors, assembled each individually, their use in instruments with much more than 100 detectors requires an enormous assembling time.  

\subsection{Detectors arrays}

There is at present strong motivation to produce bolometer arrays to expand the focal plane format and hence beat the photon noise. Bolometers using superconducting transition-edge sensors, for instance, permit to make large arrays built in a collective way by lithographic techniques. In addition, such sensors offer many advantages with respect to classical semi-conducting thermistors, such as excellent linearity or fast time response. The detector readout using SQUID (Superconducting Quantum Interference Device) via multiplexing permits to limit the number of readout channels.

\subsection{Antenna-coupled bolometers} 
In classical bolometers, the radiation power is dissipated in an absorber. A thermometer measures the temperature change of the absorber which is, to first order, proportional to the incident EM power. 

A new concept being now developed is using a lithographed antenna to convert the EM power into a current. This current can be electronically filtered and combined with other current to be thermally detected in a small bolometer. These techniques are now being developed but are not quite mature still for astrophysical observations.

This approach can be used for polarization measurements by using properly shaped antennas. An alternate interesting approach would be to combine currents in order to produce directly a signal proportional to one of the Stokes parameters. Although this scheme needs to be fully demonstrated, it is seriously considered for precise polarization measurements from Antarctica.

\section{A few existing and planned CMB polarization experiments} 

\subsection{AMIBA}

The Array for Microwave Background Anisotropy (AMiBA\footnote{http://amiba.asiaa.sinica.edu.tw/}) is being built for measurements of small--scale CMB temperature and polarization anisotropy as well as for observations of the Sunyaev-Zel'dovich Effect \cite{2000IAUS..201E..31L}. The project, led by the Academia Sinica Institute of Astronomy and Astrophysics in Taiwan, is an interferometer array with 19 elements mounted on a common platform. 
 
The array will operate around 90--100 GHz, with full polarization capabilities (including $V$) and a 20 GHz correlation bandwidth. Two sets of array dishes are planned, for two configurations. 1.2 meter dishes will be used for high resolution observations, and 0.3 meter dishes for a larger field of view. Initial observations targeting the $E$--$E$ spectrum, using a subset of the array, are scheduled for 2004.

The choice of the central observation frequency of 90 GHz, located at a minimum of foreground total emission, permits also to avoid the expected contamination by radio sources which is a permanent problem at lower frequencies. As a drawback, atmospheric emission in the frequency band requires to put the instrument in a very dry and high site. AMIBA is to be deployed on Mauna Kea, Hawaii.

\subsection{Archeops}

The Archeops experiment\footnote{http://www.archeops.org/} is a balloon--borne bolometer CMB anisotropy experiment.
A first determination of the Galactic polarized emission at 353~GHz has been obtained from Archeops data taken during the Arctic night of February 7, 2002 after the balloon--borne instrument was launched by CNES from the Swedish Esrange base near Kiruna \cite{archeops-polar}. In addition to unpolarized 143~GHz, 217~GHz and 545~GHz detectors, Archeops had six 353~GHz bolometers mounted in three polarization sensitive pairs that were used for Galactic foreground studies. Maps of the $I,~Q,~U$ Stokes parameters over 17~\% of the sky and with a 13~arcmin resolution show a significant Galactic large scale polarized emission coherent on the longitude ranges [100, 120] and [180, 200] deg. with a degree of polarization at the level of 4--5~\%, in agreement with expectations from starlight polarization measurements. Some regions in the Galactic plane (Gem~OB1, Cassiopeia) show an even stronger degree of polarization in the range 10--20~\%. Extrapolated to high Galactic latitude, these results indicate that interstellar dust polarized emission is a major foreground for CMB polarization measurement.

\subsection{B2K}

The \boom experiment, in a new ''B2K" version\footnote{http://oberon.roma1.infn.it/boomerang/b2k/} equipped with an entirely new linearly polarized receiver, successfully flew a long duration balloon flight from McMurdo Station, Antarctica during the Austral summer of 2002-3 \cite{montroy}. The focal plane consists of a set of 4 PSB pairs operating at 150 GHz, together with 4 two-color photometers sensitive to a single linear polarization.  The photometer bands are centered at 245 and 345 GHz.  \boom mapped a region of the sky centered near RA,DEC (J2000) near 80,-44.  The experiment mapped about 3\% of the sky with a sensitivity at 150 GHz of about 20 $\mu \mathrm{K}_\mathrm{CMB}$ per 10 arcminute pixel, and the central 100 deg${}^2$ of that region with an average 
senstivity of about 4.2 $\mu \mathrm{K}_\mathrm{CMB}$. Single detector NETs of $\sim 150 \mu\mathrm{K}_\mathrm{CMB}\sqrt{s}$ at 150 GHz were achieved with the PSBs.
In addition, B2K also mapped a $\sim $ 400 deg${}^2$ cut centered on the Galactic plane with a sensitivity of about 12  $\mu \mathrm{K}_\mathrm{CMB}$ per pixel at 150 GHz.  Both these regions were also observed at 245 and 345 GHz with the polarized photometers, with $\sim 7$ arcminute beams.

The B2K experiment is also a testbed to demonstrate the performance of PSBs for CMB polarization measurement in a configuration close to that of the Planck HFI. B2K data analysis is currently under way.

\subsection{BICEP}

BICEP\footnote{http://www.astro.caltech.edu/~lgg/bicep\_front.htm}
 (Background Imaging of Cosmic Extragalactic polarization) is an experiment using refractive optics and an array of 96 polarization sensitive bolometers cooled to 250mK to achieve 1$^\circ$ and 0.7$^\circ$ beams at 100 GHz and 150 GHz respectively. This experiment will be placed in the South Pole in 2004 and will map a large region of the sky around the South Celestial Pole. BICEP is complementary to the QUEST experiment (see below), which will measure smaller scales.

\subsection{CAPMAP}

CAPMAP\footnote{http://cosmology.princeton.edu/capmap/} (Cosmic Anisotropy polarization Mapper) is a new experiment using PIQUE detector technology (see below), but observing at smaller angular scales (3.6 arcminutes at 100 GHz). CAPMAP is planned to use 12 correlation polarimeters at 100 GHz and 4 correlation polarimeters at 40 GHz, located at the focus a large (seven meter diameter) off--axis telescope from Lucent Technologies in Crawford Hill, New Jersey. A prototype with 4 100--GHz polarimeters has run during the 2002 winter.

CAPMAP observes mostly at 100 GHz, where galactic foregrounds are supposed to be as low as possible. However, the location of the antenna in New--Jersey restricts the useful observing time to a small fraction of winter because of atmosphere.

\subsection{COMPASS}

The COMPASS\footnote{http://cmb.physics.wisc.edu/compass.html} (Cosmic Microwave Polarization at Small Scales) experiment is an on--axis 2.6 meter telescope coupled with a correlation polarimeter operating in the Ka band around 31 GHz with HEMT amplifiers \cite{farese-compass-inst}.

Special care has been taken to avoid systematics as those induced by mirror deformation from solar heating, or spurious polarization from struts holding the secondary. Thermal effects are avoided by insulating carefully the back of the telescope, and the secondary is supported by a radio--transparent conical support made of expanded polystyrene.

The COMPASS experiment yielded recently an upper limit of CMB polarization at 20 arcminute angular scales \cite{farese-compass-res}.

\subsection{DASI}

The DASI\footnote{http://astro.uchicago.edu/dasi/} team has been first to claim a detection of CMB $E$--type polarization, using a ground--based interferometer located at the south pole \cite{2002Natur.420..763L}. An $EE$ polarization of 0.8 $\mu$K was detected with 4.9 sigma significance using data collected at the South Pole in 2000.

The DASI interferometer\footnote{http://astro.uchicago.edu/dasi/} consists of an array of thirteen feed horns, which operates in 10 1-GHz bands over the frequency range 26--36~GHz. The interferometer samples multipoles in the
range $l\simeq$140--900. The entire receiver set is attached
to an altitude-azimuth mount, so that projected baselines do not change as a source is tracked around the sky.  
For each feed, a mechanically switchable waveguide polarizer is inserted between the amplifier and the feedhorn to select between left- and right-handed polarization states, $L$ and $R$. Polarization states are correlated to produce directly measurements of linear combinations of $T$ and $V$ and of $Q$ and $U$ in the multipole (or Fourier) space.

\subsection{MAXIPOL}

The MAXIMA experiment, which produced one of the first very clear detection of the first Doppler peak of CMB anisotropies, was converted to MAXIPOL\footnote{http://groups.physics.umn.edu/cosmology/maxipol/} by adding polarizers and a half-wave rotating plate in front of the focal plane \cite{johnson}. The rotation of the half--wave plate at 2 Hz modulates polarization at twice that frequency, which permits to reject a fraction of systematic effects as well as  low-frequency drifts. In this set--up, one of the technical issues is to avoid systematic effects at harmonics of the half--wave plate spinning frequency. 

The MAXIPOL detectors (12 detectors at 140 GHz and 4 detectors at 420 GHz), cooled to 100 mK, have sensitivities of order 100--150 $\mu$K$\sqrt{s}$ at 140 GHz and beam sizes about 10 arcminutes. Covering a few square degrees (a few hundred resolution elements) with few microkelvin sensitivity per beam can be obtained with a standard duration flight.

\subsection{PIQUE}

The Princeton I,Q,U Experiment (PIQUE) comprises a single 90~GHz 
correlation polarimeter underilluminating a 1.4~m off-axis parabola \cite{wol97}, fed with a corrugated horn antenna to provide a beamwidth of $0.23^\circ$. The instrument observes a single Stokes
parameter in a ring of radius $1^\circ$ around the north celestial pole.

The RF signals from the two arms of an OMT (oriented so one arm is parallel to the azimuthal scan direction) are mixed down to a 2--18~GHz intermediate frequency, split into three sub-bands, and then directly multiplied together in a broad bandwidth mixer. A mechanical refrigerator cools the corrugated feed horn, the orthomode transducer (OMT), and the HEMT amplifiers to $\lesssim 40$~K. 

Because the correlation polarimeter directly measures the polarized electric field, rather than detecting and then differencing two large intensity signals, the instrumental set--up is largely immune to strong leakages of $I$ into $Q$ and/or $U$. 

\subsection{Planck}

The Planck mission,\footnote{http://astro.estec.esa.nl/Planck/} to be launched by ESA in 2007, has been designed primarily for mapping CMB temperature anisotropies, with an angular resolution of about 4.5 arcminutes, and a sensitivity of $\Delta T/T ~ 2 \times 10^{-6}$ per resolution element \cite{bouchetpiatCRAS}. The original design proposed after the preliminary design study has been since then slightly modified for linear polarization measurement capability.

The Planck optics comprise a 1.5 meter  useful diameter off-axis gregorian telescope, at the focal plane of which are installed two complementary instruments, the HFI and the LFI. The HFI (High Frequency Instrument) is an array of bolometers cooled to 0.1 K, which observe the sky in six frequency channels from 100 to 850 GHz. The HFI uses PSBs to be polarization sensitive at 350, 220, 150, and possibly 100 GHz. The LFI is an array of radiometers observing the sky at 30, 44 and 70 GHz with polarization sensitivity in all channels.

Planck will observe the sky from the L2 Sun-Earth Lagrange point, in a very stable thermal environment, away from sources of spurious radiation due to the Earth, the Moon and the Sun. The scanning is made along large circles around an anti--solar spin axis at a rate of 1 rpm. The spin axis follows roughly the apparent motion of the Sun, so that the full sky is covered in slightly more than 6 months.

The Planck polarization--measurement set--up takes advantage of the possibility to measure two orthogonal polarizations using the same feed horn, by splitting the orthogonal polarizations into two different detectors.
For each such feed, a pair of detectors sensitive to orthogonal polarizations (polarimeters) share the same optics (filters, waveguides, corrugated horns, and telescope), but have different readouts. One single horn produces two signals corresponding to measurements integrated in principle over the same beam shape, but with orthogonal linear polarizations.

The sensitivity of Planck to polarization, of the order of ten microkelvin per resolution element on average and about ten times more in highly redundant patches, will permit to put strong constrains on the polarization power spectra of the CMB.

\subsection{POLAR}

POLAR\footnote{http://cmb.physics.wisc.edu/polar/} (Polarization Observations of Large Angular Regions) is a first generation CMB polarization experiment which was designed to look for sky polarization at large angular scales, in the Ka frequency band (26-36 GHz) \cite{2003ApJS..144....1K}. POLAR is a single--pixel correlation polarimeter which collected about 750 hours of data in 2000. Being single--pixel, it is far from being sensitive enough for detecting the very low polarization modes on large scales. For this reason, the POLAR experiment has been stopped in favor of COMPASS, after producing an upper limit of CMB polarization at low $\ell$ \cite{2001ApJ...560L...1K}.

\subsection{PolarBear}

The PolarBear experiement (POLARization of the Background millimEter bAckground Radiation) is a CMB polarization experiment dedicated to characterizing the E-modes and search for B-modes signature of the CMB. PolarBear will use an antenna coupled TES bolometer array cooled to 300 mK. Such technology allows having a focal plane with several hundred of detectors at different frequencies: for PolarBear, it is planned to have 5 different frequency bands between 90 GHz and 350 GHz, with 150 or 300 pixels in each band, which offers significant frequency coverage of prime importance for foreground monitoring and subtraction. This instrument will be placed in the focal plane of a 3m telescope located in White Mountain (CA) and the first light is expected in 2005. The target sensitivity is of 1.5 $\mu$K RMS per 5 arcminute pixel on a 15 degree by 15 degree CMB polarization map.

\subsection{QUEST/QUaD}

QUEST\footnote{http://www.stanford.edu/group/quest\_telescope/quest\_web/quest\_instrument.htm} \cite{2002apb..conf..159P} is a bolometer-based polarimeter designed for CMB polarization measurements.
The QUEST focal plane consists of 12 PSB pairs at 100 GHz and 19 pairs at 150 GHz. The expected sensitivity is about 300 $\mu$K$_{\rm CMB}\sqrt{s}$ at 150 GHz. 

A dedicated 2.6 meter Cassegrain telescope, mounted on the former DASI mount, will provide a resolution of 4 arcminutes at 150 GHz. The assembly of QUEST on the DASI mount (QUEST and DASI) is named QUaD.

The steerable DASI mount can also be rotated along the line of sight, which can be used along with a rotating half-wave
plate to modulate the polarization response for rejecting systematics. The half-wave plate will be not be synchronously rotated as for MAXIPOL, but rather set at a fixed angle for each scan, which is technologically easier and avoids vibrations which can be a source of microphonics in bolometer instruments.

\subsection{SPORT}

SPOrt\footnote{http://sp0rt.bo.iasf.cnr.it:8080/Docs/Public/Project/the\_project.php} is a large angular scale radiometer experiment, selected by ESA to be embarked on the International Space Station (ISS) for a minimum lifetime of 18 months starting (in principle) in 2005, although perturbations in the manned space mission flights may impact the original calendar. SPOrt is designed to perform direct measurements of the $Q$ and $U$ Stokes parameters in three frequency channels (22, 32 and 90 GHz), which permits to plan CMB measurements at 90 GHz while monitoring low-frequency galactic foregrounds at 22 and 33. The detectors are correlation polarimeters using Ortho-Mode Transducers (OMT) to split the incoming radiation into two circularly polarized waves.

The SPOrt experiment is expected to map about 80\% of the sky, with a scanning strategy imposed by constraints due to its location on the ISS.

The SPOrt resolution will be about 7$^\circ$, and the sky coverage of about 80\%, for a final expected sensitivity of about 2 $\mu$K per resolution element for a 18--month mission.

\subsection{WMAP}

WMAP\footnote{http://map.gsfc.nasa.gov/} is a space mission primarily designed to be the second generation spaceborne CMB anisotropy experiment after \emph{COBE}. WMAP detectors, however, are polarization sensitive. WMAP is at present the most sensitive operating CMB experiment, essentially thanks to the very stable environment at the Sun-Earth L2 Lagrange point, and to the large integration time of spaceborne instruments.

The first analysis of WMAP data, including polarization, has permitted to measure the $E$--$T$ power spectrum of the CMB with unprecedented accuracy, confirming theoretical expectations for primary fluctuations, while detecting a large bump on the large scale cross power spectrum, interpreted as due to early reionization \cite{2003ApJS..148..161K}.

Unfortunately for polarization mapping, the WMAP experiment observes with a differential observation scheme which, while quite useful originally to reject low frequency noise in the detector readouts has since then been demonstrated to be unnecessary for future space missions as Planck thanks to new developments in map--making techniques for large size data sets \cite{1998A&AS..127..555D,2000A&AS..142..499R,2001A&A...374..358D}. This differencing scheme complicates the analysis of the data to obtain polarization maps, introducing yet another difference between large intensity signals in the polarization data. Refined data analysis is currently under way.

\section{Prospective: the Inflation Probe}

As most of the CMB community agrees to consider polarization measurement as the next challenge of CMB science, and an unique way to put constraints on inflationary models in the next ten years, the idea
to build a next generation CMB experiment after Planck, dedicated to measuring CMB polarization with ultimate accuracy, has been pushed forward by a large fraction of the CMB community.

NASA, in its newest Structure and Evolution of the Universe program called {\emph{Beyond Einstein}}, has solicited proposals for concept studies of a moderate-sized space mission called Inflation Probe, the objective of which is to search for the imprint of gravitational waves from inflation in the polarization of the cosmic microwave background.
This probe is expected to map all the modes of polarization of the CMB to determine the source of this polarization on all scales, and to search the CMB specifically for the signature of gravitational waves from the Big-Bang to test theories of the very early Universe.

Such a mission should ideally be launched a few years after Planck. Its design should not be frozen before first polarization results obtained from ground-based and balloon-borne experiments clarify the various trade-offs. In particular, the final design will depend on the status of polarized foregrounds, on technology readiness, and (for complementarity with ground--based observations) on the exact impact of the atmosphere and of terrestrial environment on polarization measurements from ground.

The extreme sensitivity required for the inflation probe requires a large array of photon-noise limited detectors. For this purpose, antenna-coupled large-size bolometer arrays seem particularly promising, although the design and space qualification of such instruments will still require considerable work in the next few years.

\section{Conclusion}

The present status and prospective for experimental studies of CMB polarization have been reviewed. Tremendous activity in the field is currently underway, leading to major technological developments, increasing experience, and ever--increasing instrument sensitivity.

As mentioned above, first detections of polarization and polarization--temperature correlations have already been announced: a detection of $E$ type polarization and $T-E$ correlation by DASI, the $T-E$ detection by WMAP. It is quite likely that in the very near future, additional measurements will confirm and refine these measurements, in particular B2K and Maxipol, who have gathered data already. The Planck mission will measure the $EE$ power spectrum quite accurately on all scales.

There is, however, still a lot to be understood for measuring the very tiny expected $B$--type polarization that would permit to constrain the physics of inflation. Foregrounds and instrumental systematics, in particular, are still far from being under control, and it is quite likely that yet more than a decade of hard work, of trials and errors, will pass before the technology and methodology, both for instrument developments as for data reduction and analysis, are mature enough for closing this new chapter of CMB studies.

\section{Acknowledgements}
We acknowledge gratefully the help of Bill Jones, who wrote the  paragraph on polarization sensitive bolometers and provided useful information about the B2K experiment. JD thanks Fran\c{c}ois--Xavier D\'esert for pushing him into writing this review, which would never have happened without loads of (mailbox--cumbersome) patient reminding emails.



%

\begin{thebibliography}{99}
\selectlanguage{english}

\bibitem[Bennett et al.(2003)]{2003ApJ...583....1B} Bennett, C.~L.~et al.\ 2003, ApJ, 583, 1 


\bibitem[Beno\^{\i}t et al.(2000)]{2000A&AS..141..523B} Beno\^{\i}t, A.~et al.\ 2000, A\&A Suppl. Ser., 141, 523 

\bibitem[Beno\^{\i}t et al.(2002a)]{2002APh....17..101B} Beno\^{\i}t, A. et al., 2002, Astroparticle Physics, 17, 101

\bibitem[Beno\^{\i}t et al.(2002b)]{archeops-polar} Beno\^{\i}t, A. et al., 2003, to appear in A\&A (astro-ph/0306222)

\bibitem[Bouchet et al.(2003)]{bouchetpiatCRAS}Bouchet, F.R., Piat, M., and Lamarre, J.--M.,\ 2003, In {\em The Cosmic Microwave Background: present status and
cosmological perspectives}, This issue of C. R. (Physique). Academie des Sciences, Paris

\bibitem[Carretti et al.(2003)]{2003SPIE.4843..305C} Carretti, E.~et al.\ 
2003, Proceedings of the SPIE, 4843, 305 

\bibitem[Church et al.(1997)]{1997ApJ...484..523C} Church, S.~E., Ganga, 
K.~M., Ade, P.~A.~R., Holzapfel, W.~L., Mauskopf, P.~D., Wilbanks, T.~M., 
\& Lange, A.~E.\ 1997, ApJ, 484, 523 
\bibitem[Couchot et al.(1999)]{1999A&AS..135..579C} Couchot, F., Delabrouille, J., Kaplan, J., \& Revenu, B.\ 1999, A\&A Suppl. Ser., 135, 579 

\bibitem[Delabrouille(1998)]{1998A&AS..127..555D} Delabrouille, J.\ 1998, 
A\&A Suppl. Ser., 127, 555 
\bibitem[Delabrouille, J.(2003)]{Planck-polar} Delabrouille, J. (astro-ph/0307550)

\bibitem[De Petris et al.(1999)]{1999NewAR..43..297D} De Petris, M.~et al.\ 
1999, New Astronomy Review, 43, 297 

\bibitem[De Petris et al.(2002)]{2002ApJ...574L.119D} De Petris, M.~et al.\ 
2002, ApJ letters, 574, L119 

\bibitem[D{\'e}sert et al.(2002)]{2002ecmw.conf..116D} D{\'e}sert, 
F.-X.~et al.\ 2002, AIP Conf.~Proc.~616: Experimental Cosmology at 
Millimetre Wavelengths, 116 

\bibitem[Dor{\'e} et al.(2001)]{2001A&A...374..358D} Dor{\'e}, O., 
Teyssier, R., Bouchet, F.~R., Vibert, D., \& Prunet, S.\ 2001, A\&A, 374, 358 
\bibitem[Farese et al.(2003a)]{farese-compass-inst} Farese, P.C.~et 
al.\ 2003, To be published in the proceedings of "The Cosmic Microwave Background and its Polarization", New Astronomy Reviews, S. Hanany and K.A. Olive eds. (astro-ph/0305608) 

\bibitem[Farese et al.(2003b)]{farese-compass-res} Farese, P.C. et al., 2003, submitted to ApJ (astro-ph/0308309)

\bibitem[Grainge et al.(2003)]{2003MNRAS.341L..23G} Grainge, K.~et al.\ 
2003, MNRAS, 341, L23 

\bibitem[Hanany et al.(2000)]{2000ApJ...545L...5H} Hanany, S.~et al.\ 2000, 
ApJ letters, 545, L5 

\bibitem[Holzapfel et al.(1997)]{1997ApJ...479...17H} Holzapfel, W.~L., 
Wilbanks, T.~M., Ade, P.~A.~R., Church, S.~E., Fischer, M.~L., Mauskopf, 
P.~D., Osgood, D.~E., \& Lange, A.~E.\ 1997, ApJ, 479, 17 

\bibitem[Johnson et al.(2003)]{johnson} Johnson, B.R.~et 
al.\ 2003, To be published in the proceedings of "The Cosmic Microwave Background and its Polarization", New Astronomy Reviews, S. Hanany and K.A. Olive eds. (astro-ph/0308259) 

\bibitem[Jones et al.(2003)]{jones-psb} Jones, W. et al., 2003, Proceedings of the SPIE, 4855, 227, Millimeter and Submillimeter Detectors for Astronomy; Thomas G. Phillips, Jonas Zmuidzinas eds. (astro-ph/0209132)

\bibitem[Kaplan et al., 2002]{kaplanCRAS} Kaplan, J.~et 
al.\ 2003, In {\em The Cosmic Microwave Background: present status and
cosmological perspectives}, This issue of C. R. (Physique). Academie des Sciences, Paris

\bibitem[Keating et al.(2001)]{2001ApJ...560L...1K} Keating, B.~G., O'Dell, 
C.~W., de Oliveira-Costa, A., Klawikowski, S., Stebor, N., Piccirillo, L., 
Tegmark, M., \& Timbie, P.~T.\ 2001, ApJ Letters, 560, L1 

\bibitem[Keating et al.(2003)]{2003ApJS..144....1K} Keating, B.~G., O'Dell, 
C.~W., Gundersen, J.~O., Piccirillo, L., Stebor, N.~C., \& Timbie, P.~T.\ 
2003, ApJ Suppl. Ser., 144, 1 

\bibitem[Kogut et al.(2003)]{2003ApJS..148..161K} Kogut, A.~et al.\ 2003, 
ApJ Suppl. Ser., 148, 161 
\bibitem[Krauss(1986)]{krauss} Krauss, J. D., 1986, Radio
Astronomy, (2d ed.; Powell, OH: Cygnus-Quasar Books)

\bibitem[Kuo et al.(2003)]{kuo-acbar} Kuo, C. L. et al., 2003, submitted to ApJ (astro-ph/0212289) 

\bibitem[Leitch et al.(2002a)]{2002ApJ...568...28L} Leitch, E.~M.~et al.\ 
2002, ApJ, 568, 28 

\bibitem[Leitch et al.(2002b)]{2002Natur.420..763L} Leitch, E.~M.~et al.\ 
2002, Nature, 420, 763 

\bibitem[Lo et al.(2000)]{2000IAUS..201E..31L} Lo, K.~Y., Chiueh, T., 
Liang, H., Ma, C.~P., Martin, R., Ng, K.-W., Pen, U.~L., \& Subramanyan, 
R.\ 2000, IAU Symposium, 201,  

\bibitem[Mauskopf et al.(2000)]{2000ApJ...538..505M} Mauskopf, P.~D.~et 
al.\ 2000, ApJ, 538, 505 

\bibitem[Montroy et al.(2003)]{montroy} Montroy, T.~et 
al.\ 2003, To be published in the proceedings of "The Cosmic Microwave Background and its Polarization", New Astronomy Reviews, S. Hanany and K.A. Olive eds. (astro-ph/0209132) 

\bibitem[Pearson et al.(2003)]{2003ApJ...591..556P} Pearson, T.~J.~et al.\ 
2003, ApJ, 591, 556 
\bibitem[Piacentini et al., 2002]{2002ApJS..138..315P} Piacentini, F.~et 
al.\ 2002, ApJ Suppl. Ser., 138, 315

\bibitem[Piccirillo et al.(2002)]{2002apb..conf..159P} Piccirillo, L.~et 
al.\ 2002, AIP Conf.~Proc.~609: Astrophysical Polarized Backgrounds, 159 

\bibitem[Pointecouteau et al.(1999)]{1999ApJ...519L.115P} Pointecouteau, 
E., Giard, M., Benoit, A., D{\'e}sert, F.~X., Aghanim, N., Coron, N., 
Lamarre, J.~M., \& Delabrouille, J.\ 1999, ApJ letters, 519, L115 

\bibitem[Revenu et al.(2000)]{2000A&AS..142..499R} Revenu, B., Kim, A., 
Ansari, R., Couchot, F., Delabrouille, J., \& Kaplan, J.\ 2000, A\&A Suppl. Ser., 142, 499 

\bibitem[Smoot et al.(1990)]{1990ApJ...360..685S} Smoot, G.~et al.\ 1990, ApJ, 360, 685 
\bibitem[Wollack et al., 1997]{wol97} Wollack, E.J., 
Devlin, M.J., Jarosik, N., Netterfield, C. B., Page, L., \&
Wilkinson, D. 1997, ApJ, 476, 440

\end{thebibliography}
\end{document}